\documentclass[aip,twocolumn,reprint]{revtex4-1}
\usepackage{graphicx}
\usepackage{amsmath,amsthm}

\draft % marks overfull lines with a black rule on the right

\begin{document}

%\preprint{}

\title{A parallel algorithm for implicit depletant simulations} %Title of paper



\author{Jens Glaser}
%\thanks{}
\author{Andrew S. Karas}
\affiliation{Department of Chemical Engineering, University of Michigan, 2800 Plymouth Rd. Ann Arbor, MI 48109, USA}
\author{Sharon C. Glotzer}
\affiliation{Department of Chemical Engineering, University of Michigan, 2800 Plymouth Rd. Ann Arbor, MI 48109, USA}
\affiliation{Department of Materials Science and Engineering, University of Michigan, 2300 Hayward St. Ann Arbor, MI 48109, USA}
\email{sglotzer@umich.edu}


\date{\today}

\begin{abstract}
We present an algorithm to simulate the many-body depletion interaction between
anisotropic colloids in an implicit way, integrating out the degrees of freedom
of the depletants, which we treat as an ideal gas. Because the depletant
particles are statistically independent and the depletion interaction is
short-ranged, depletants are randomly inserted in parallel into the excluded
volume surrounding a single translated and/or rotated colloid. A
configurational bias scheme is used to enhance the acceptance rate. The method
is validated and benchmarked both on multi-core CPUs and graphics processing
units (GPUs) for the case of hard spheres, hemispheres and discoids. With
depletants, we report novel cluster phases, in which hemispheres first assemble
into spheres, which then form ordered hcp/fcc lattices. The method is
significantly faster than any method without cluster moves and that tracks
depletants explicitly, for systems of colloid packing fraction $\phi_c<0.50$,
and additionally enables simulation of the fluid-solid transition.
\end{abstract}


\maketitle %\maketitle must follow title, authors, abstract and \pacs

\section{Introduction}

The self-assembly of anisotropic particles into complex structures has emerged
as a promising strategy towards the fabrication of materials with novel
properties \cite{Glotzer2007}.  Methods for the synthesis of anisotropic nano-
and colloidal particles \cite{Sacanna2013,Xia2015} are becoming available, and
enable experiments that study their phase behavior
\cite{Sacanna2010,Henzie2012,Ye2013}. Anisotropic particles, such as proteins,
are also emerging building blocks for biomaterials
\cite{Liljestrom2014,Park2014}.  Simulations predict a wealth of different
crystal structures that hard shapes form through maximization of entropy.
\cite{Damasceno2012,Agarwal2011} In addition to particle shape, attractive
interactions between patchy particles can be important in achieving desired
target structures \cite{Tang2006,Ye2013,Ye2013c}. Towards that end, the main
routes that are actively being explored include surface functionalization of
nanoparticles using short DNA molecules \cite{Jones2010,Auyeung2014}, and
exploiting the depletion interaction between colloids in the presence of small
polymer chains \cite{Sacanna2010, Rossi2011, Rossi2015}. Here we focus on the
depletion interaction, since it is of entropic origin and arises without the
need for engineering particle surface chemistry, emerging in mixtures of
colloids with non-adsorbing polymer.

Depletion\cite{Asakura1954a} describes the emergent attraction between colloids
in solution that maximize the free volume available to a small-particle
cosolute via overlap of their excluded volume shells.  It has been demonstrated
that depletion enhances the directional entropic forces
\cite{Damasceno2012,Young2013a,Anders2014,Anders2014a} resulting from anisotropic particle
shape, and that it promotes the contact between large facets. The depletion
interaction can promote binding between lock and key colloids
\cite{Sacanna2010,Colon-Melendez2015} and lead to the formation of porous
phases \cite{Ashton2015}.  Because depletion mediates an additional attraction
of entropic origin, this interaction can be thought of as competing with
contact (excluded volume) interactions resulting from particle shape. Depletion
thus enables novel phase behavior through the additional parameters of
depletant shape and density \cite{Rossi2015,Karas2015a}.  Therefore, it is
desirable to have a method to investigate the self-assembly of anisotropic
shapes in the presence of depletants. Results for the phase behavior of binary
hard sphere mixtures have been reported\cite{Dijkstra1998} using thermodynamic
integration. In general, however, such results are challenging to obtain
because of the size disparity between the colloid and the depletant. If one is
interested in the phase behavior of the colloids, a customary approximation
treats the depletant particles as an ideal gas\cite{Asakura1954a,Widom1970}.
This approximation would, in principle, allow integrating out the depletant to
arrive at an effective colloid-colloid interaction; however, the resulting
interaction is a many-body interaction and we are not aware of any prior
implementation that treats many-body effects exactly. Here, we propose a novel,
parallel Monte Carlo algorithm to simulate the depletion interaction between
arbitrarily shaped colloids in an efficient manner that includes many-body
effects.

\begin{figure}
\centering\includegraphics[width=\columnwidth]{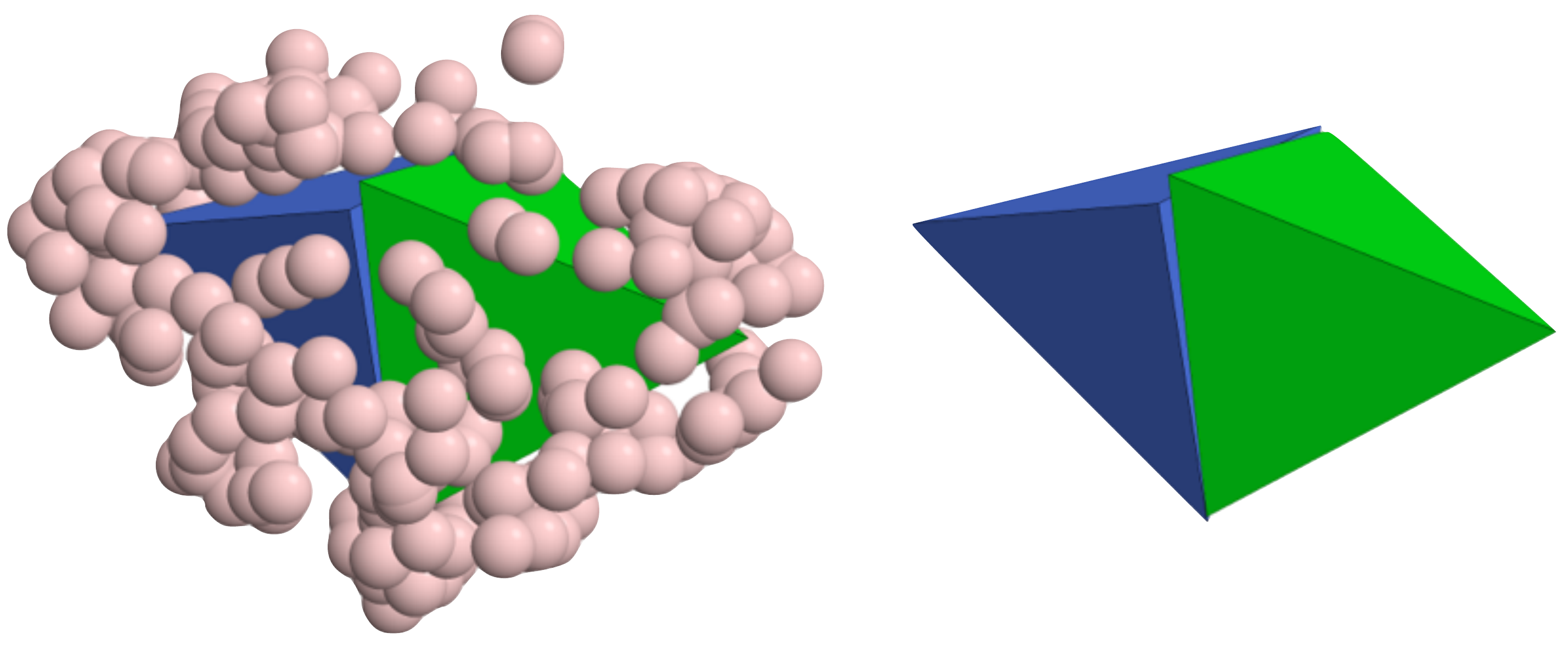}
\caption{Explicit ({\em left}) vs. implicit ({\em right}) treatment of depletion
interactions. Hard tetrahedra in solution with small, penetrable hard spheres
aggregate face to face, to maximize the free volume available to the depletants.}

\label{fig:explicit_implicit}
\end{figure}

Figure \ref{fig:explicit_implicit} shows the effect of depletion interactions
between two hard tetrahedra in solution with small penetrable hard
spheres.  The small spheres mediate an attractive interaction between the
colloids that drives them to aggregate face to face.  For two particles only,
the depletion interaction can be easily simulated explicitly (left panel)
or implicitly (right panel).  However, implicit simulation of depletion
interactions allow for a tremendous performance benefit, particularly for
dilute systems of colloids and high densities of depletants, as we
demonstrate below.

This paper is organized as follows. In section \ref{sec:background}, we discuss
previous numerical methods for the simulation of depletion interactions.  We
describe our algorithm in section \ref{sec:algorithm}, and validate it against
published data for hard spheres in the following section ~\ref{sec:validation}.
Section \ref{sec:results} contains new results for hemispheres and
discoids\cite{Hsiao2015a} in the presence of depletants, obtained with the new
algorithm. Finally, in Sec.~\ref{sec:conclusion} we summarize and give an
outlook on future applications of the method.

\section{Background}
\label{sec:background}
Previous numerical treatments of depletion interactions employ cluster moves.
Biben, Bolhuis and Frenkel proposed a configurational bias approach
\cite{Bolhuis1994,Biben1996}, where depletants overlapping with a moved colloid
are reinserted to enhance the acceptance probability of colloid moves.  A
geometric cluster algorithm has also been proposed by Dress and Krauth
\cite{Dress1995}, which is rejection-free and can therefore greatly enhance the
equilibration of dilute systems of colloids.  However, when the system is dense
in colloids, clusters can span the system and the algorithm ceases to be
efficient \cite{Ashton2013c}.  To explore the phase behavior of a system of
hard spheres in penetrable hard-sphere depletants, Vink and Horbach proposed
grand-canonical simulation of both the colloids and the depletants, and they
could efficiently sample the gas-liquid coexistence curve \cite{Vink2004}.
However, their scheme does not generalize well beyond to the fluid-solid
transition, because it is based on particle insertion.

All these methods have in common that they track the small depletant
particles explicitly, which are stored in memory. An interesting alternative
was proposed by Dijkstra et al.~\cite{Dijkstra2006}, who proposed a Monte
Carlo integration of the free volume around every single moved colloid.
However, their scheme does not obey detailed balance, and achieving sufficient
accuracy comes at the expense of computation time, as we discuss in more detail
below. Another implicit implementation of the depletion interaction between
octahedra was proposed by Henzie et al.~\cite{Henzie2012}, where the
generally anisotropic many-body interaction is reduced to an isotropic pair
potential.  We note that such a drastic simplification, while rendering the
problem computationally tractable, is insufficient to allow the study of
arbitrary shapes.

The scheme we describe in the following section is a completely general
treatment of depletion interactions between anisotropic particles due to an
ideal gas of depletants, and works well both for dilute and dense systems. In
the ideal gas treatment, depletants interact with colloids but not with each
other.  The algorithm is rigorous, i.e.\ it obeys detailed balance, and it can
be efficiently implemented on multi-core processors and graphics processing
units (GPUs).

\section{Description of the algorithm}
\label{sec:algorithm}

\subsection{Semigrand N$\mu_p$VT ensemble}
We simulate a semigrand ensemble of $N$ colloids in a grand-canonical bath of
penetrable depletants of chemical potential $\mu_p$.
The partition sum for the depletants is
\begin{eqnarray}
e^{-\beta\Xi{\{\vec r_{c,i}\}}}
&=& \sum\limits_{N_p = 0}^\infty \frac{e^{\beta \mu_p N_p}}{N_p! \lambda_p^{3 N_p}} \int d\vec r^{N_p}_{p,i}
e^{-\beta H_{cc}-\beta H_{cp}}\\
&=&\sum\limits_{N_p = 0}^\infty \frac{e^{\beta \mu_p N_p}}{N_p! \lambda_p^{3 N_p}} \int d\vec r^{N_p}_{p,i}
e^{-\beta H_{cc}} V_f^{N_p}
\end{eqnarray}
where $V_f = V_f[\vec r_{c,i}]$ is the free volume available to depletants and
$\lambda_p$ the thermal de Broglie wavelength associated with the depletants.
We denote the colloid-colloid contribution to the Hamiltonian as $H_{cc} =
\sum_{i,j \in \mathrm{colloids}} U_{ij}$, where $U_{ij} = \infty$ for two
colloids that overlap, and $U_{ij} = 0$ otherwise. The colloid-polymer
contribution to the Hamiltonian $H_{cp}$ is defined analogously.  Summation
over the number $N_p$ of depletants in the system results in
\begin{equation}
e^{-\beta\Xi{\{\vec r_{c,i}\}}}=e^{z_p V_f - \beta H_{cc}},
\label{eq:gc_ensemble}
\end{equation}
where $z_p \equiv \frac{e^{\beta \mu_p}}{\lambda_p^3}$ is the depletant fugacity.

\subsection{Basic idea}
\label{sec:basic}

Our central algorithmic result is the following Monte Carlo scheme to integrate the colloids
under the action of the effective potential $H_{\mathrm{eff}}\equiv - \beta^{-1} z_p V_f
[\vec r_{c,i}]$ occurring in Eq.~\eqref{eq:gc_ensemble}. The basic idea of the algorithm, which we present here,
is very simple, and we describe optimized versions of it in ensuing sections.

\label{sec:integration_scheme}
\begin{enumerate}
\item Propose a trial move for the colloids $M\to M'$.
\item Generate $N_p$ random depletant positions $\vec r_i^{(p)}$ uniformly in
the free volume of the old configuration $M$, where $N_p$ is chosen according
to $P_{z_p V_f}(N_p) \sim \mbox{Poisson}(V_f z_p)$, where
$\mbox{Poisson}(\lambda)$ is the Poisson distribution of mean and variance
$\lambda$. One possibility is to use rejection sampling in a larger volume $V_0
\supset V_f$.
\item Reject the trial move if any depletant overlaps with the new colloid
configuration $M'$, otherwise accept.
\end{enumerate}

In other words, we have an {\em a priori} move generation probability
\begin{eqnarray}
\label{eq:move_gen}P^{(N_p)}_{\mathrm{\tiny trial}}(M\to M') &=&P^{\mbox{\tiny coll}}_{\mbox{\tiny trial}}(M\to M')  P_{z_p V_f}(N_p)\\
\nonumber &=&P^{\mbox{\tiny coll}}_{\mbox{\tiny trial}}(M\to M') \frac{(z_p V_f)^{N_p}}{N_p!} e^{-z_p V_f},
\end{eqnarray}
where $P^{\mbox{\tiny coll}}_{\mbox{\tiny trial}}(M\to M')$ is symmetric in
$\Delta \vec r_{c,i} \leftrightarrow -\Delta \vec r_{c,i}$. In Eq.~\eqref{eq:move_gen},
we have used the definition of the
Poisson distribution $P_{z_p V_f}(N_p)$ with average $z_p V_f$, the number of depletants in the free volume.
We impose the following acceptance probability
\begin{equation}
P^{(N_p)}_{\mathrm{acc}} (M\to M') = \mathrm{min}(1,e^{-\beta\Delta H_{cc}}) e^{-\beta H_{cp}^{'(N_p)}}.
\label{eq:pacc}
\end{equation}

\begin{figure}
\includegraphics[width=\columnwidth]{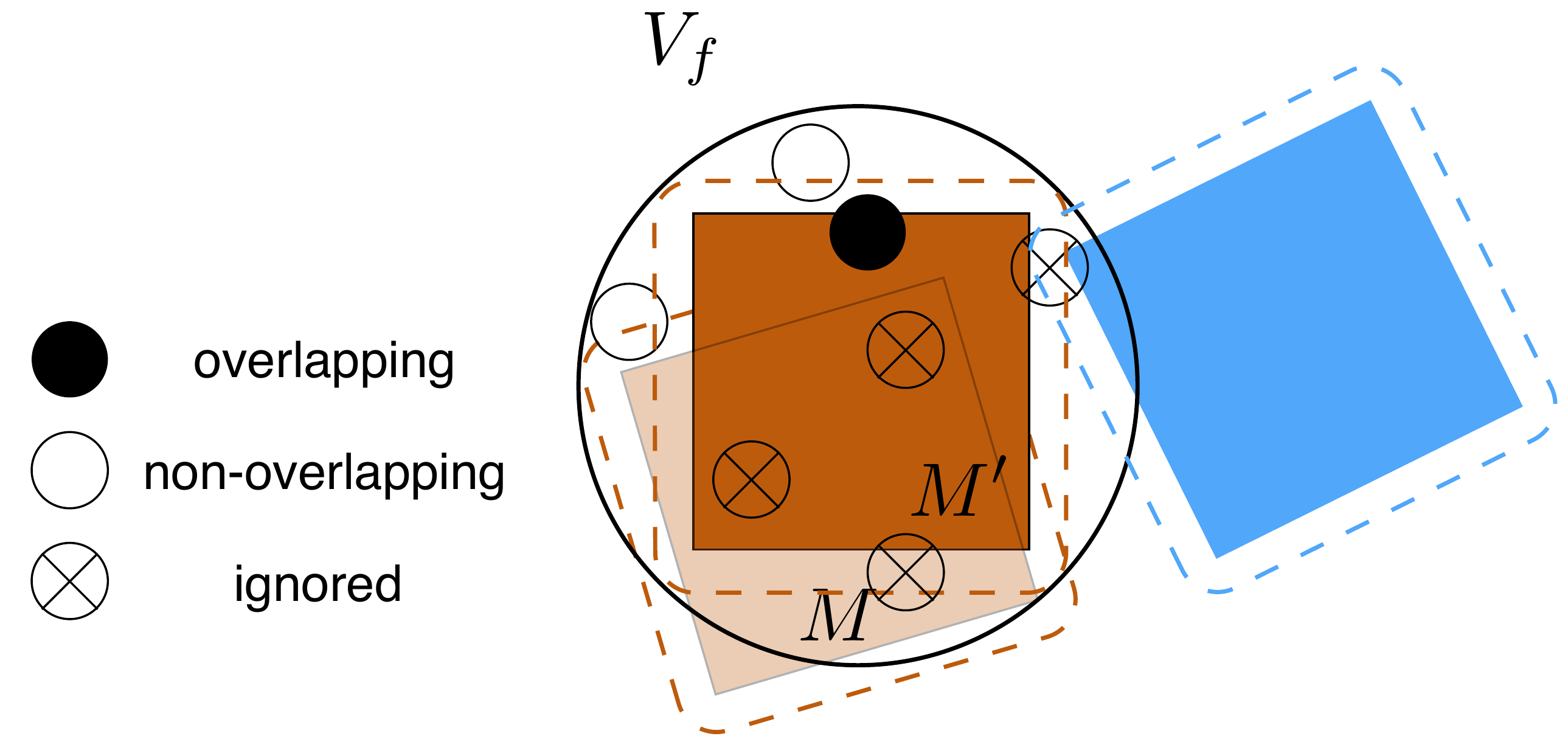}
\caption{Depletant positions (disks) considered for rejection of a colloid move (shaded
squares).  The difference between configurations $M$ and $M'$ is the position of the dark shaded
colloid. When moving the colloid to the new position, depletants are randomly inserted
into the circumsphere of the excluded volume, and depletants that only overlap with the
shape in the new configuration $M'$ lead to rejection. Depletants that overlap with
the colloid in the old position or with surrounding colloids (light shaded square) are
not considered.}
\label{fig:algorithm}
\end{figure}

Figure~\ref{fig:algorithm} contains a graphical summary of the algorithm. Here,
a square colloid is moved from configuration $M$ to configuration $M'$, by
some translation and/or rotation, and depletants are placed in the free volume.
As we detail below in Sec.~\ref{sec:optimized_algorithm}, the sampling can be
restricted to the circle (or sphere, in three dimensions) containing the
colloid in the new colloid position. By using rejection sampling, any
depletants falling into the excluded volume at the old position are ignored.
Depletants that overlap {\em only} in the new configuration lead to a rejection
of the colloid move.

Next, we show that the above scheme obeys detailed balance, which is required
for correctly sampling the ensemble defined by Eq.~\eqref{eq:gc_ensemble} in
the statistical sense. The transition probability $\pi$ from the old configuration
$M$ to the new configuration $M'$ obeys
\begin{eqnarray}
\label{eq:transition_probability}
\pi_{M\to M'}&=&e^{-\beta \Xi\{\vec r_{c,i}\}} P^{(N_p)}_{\mbox{\tiny trial}}(M\to M') P_{\mathrm{acc}}^{(N_p)}(M\to M')\nonumber\\
\nonumber&=&e^{-\beta H_{cc}+z_p V_f} P^{(N_p)}_{\mbox{\tiny trial}}(M\to M') \frac{(z_p V_f)^{N_p}}{N_p!} \\
&&\times e^{-z_p V_f}\,\mathrm{min}(1,e^{-\beta \Delta H_{cc}}) e^{-\beta H_{cp}^{'(N_p)}}
\end{eqnarray}

We require for detailed balance that $\pi_{M\to M'}=\pi_{M'\to M}$, and average over
all realizations $\left(N_p, \{\vec r_{p,i}^{N_p}\}\right)$ of depletants, in the free volume $V_f$,

\begin{eqnarray}
\label{eq:avg_pi}\sum\limits_{N_p=0}^\infty \int_{V_f} \frac{d\vec r^{N}_{p,i}}{V_f^{N_p}} \pi_{M\to M'}
&=&e^{-\beta H_{cc}} P^{\mbox{\tiny coll}}_{\mbox{\tiny trial}}(M\to M')\\
\nonumber&&\times \mathrm{min}(1,e^{-\beta \Delta H_{cc}})\\
&&\times \sum\limits_{N_p=0}^{\infty}\, \frac{(z_p V_f)^{N_p}}{N_p!} \int_{V_f}\frac{d\vec r^{N_p}_{p,i}}{V_f^{N_p}}
e^{-\beta H_{cp}^{'(N_p)}}.\nonumber
\end{eqnarray}

Note that in order to obtain Eq.~\eqref{eq:avg_pi}, we observe that the Poisson
distribution is normalized in such a way so as to cancel out the depletant
contribution, $e^{V_f z_p}$ to the ensemble weight.
The integrand in the last line of Eq.~\eqref{eq:avg_pi} is non-zero exactly for $\vec r_{p,_i} \in V_f'$;
hence, after performing the summation over $N_p$, the transition probability becomes
\begin{eqnarray}
\label{eq:detailed_bal_product}
\pi_{M \to M'} &=& e^{-\beta H_{cc}} P^{\mbox{\tiny coll}}_{\mbox{\tiny trial}}(M\to M')\, \mathrm{min}(1,e^{-\beta \Delta H_{cc}})\nonumber
\\
&& e^{z_p \mu(V_f \cap V_f')},
\end{eqnarray}
where the volume $\mu(V_f \cap V_f')$ is the intersection of the free volume
$V_f$ in the old configuration and the free volume $V_f'$ in the new
configuration. This term arises because of the integration domain in
Eq.~\eqref{eq:avg_pi}. Because of the symmetry of the Metropolis criterion,
\begin{equation}
e^{-\beta H_{cc}} \mathrm{min(1,e^{-\beta \Delta H_{cc}})} = e^{-\beta H_{cc}'} \mathrm{min}(e^{-\beta\Delta H'_{cc}},1)
\end{equation}
and the symmetry property of the set intersection, the product in Eq.~\eqref{eq:detailed_bal_product} is
symmetric under the exchange $M\leftrightarrow M'$. Consequently, our integration scheme obeys detailed balance.

\subsection{Improved formulation}
\label{sec:optimized_algorithm}
The above integration scheme conveys the general idea of the algorithm.
However, this algorithm is impractical to implement as is in an actual program,
because it would require computation of the free volume $V_f$ in the entire
simulation box for every single colloid move. Without loss of generality, we
can restrict the sampling volume $V_f$ for depletants to a smaller volume $V_0
\supseteq V_{\mathrm{excl}}' \ \backslash V_{\mathrm{excl}}$, i.e.\ containing
the excluded volume $V_{\mathrm{excl}}^{'}$ of the colloids in the system in
the new configuration minus the excluded volume $V_{\mathrm{excl}}$ in the old
configuration.  The improved scheme is the same as the old scheme
(Sec.~\ref{sec:basic}), as are the move generation and acceptance
probabilities, with the exception that $V_f$ is replaced by $V_f \cap V_0$.
The proof of detailed balance is only slightly more complicated for this
algorithm.

We rewrite the ensemble weight
\begin{eqnarray}
\nonumber\Pi_{\{\vec r_{c,i}\}}&=& e^{-\beta H_{cc}-\beta H_{\mathrm{eff}}}\\
\nonumber&=& e^{-\beta H_{cc}+z_p V_f}\\
&=&e^{-\beta H_{cc} + z_p \left[\mu(V_f \cap V_0) + \mu (V_f \cap \overline{V_0})\right]},
\label{eq:ensemble_weight_V0}
\end{eqnarray}
where $\overline{V_0}$ denotes the complement $V\backslash V_0$ with respect to the simulation volume $V$.
Using Eq.~\eqref{eq:ensemble_weight_V0}, integrating over $V_0\cap V_f$ and using transformations analogous
to Eqs.~\eqref{eq:transition_probability}-\eqref{eq:detailed_bal_product},
the transition probability $M\to M'$ averaged over the number of test depletants and their positions becomes
\begin{eqnarray}
\nonumber
\pi_{M\to M'}&=&e^{-\beta H_{cc}}\mathrm{min}\left(1,e^{-\beta \Delta H_{cc}}\right) P^{\mbox{\tiny coll}}_{\mbox{\tiny trial}}(M\to M')\\
&&e^{z_p \left[\mu(V_f\cap V_0 \cap V_f')+\mu(V_f\cap\overline{V_0})\right]}.
\label{eq:transition_prob_V0}
\end{eqnarray}

It is straightforward to show that this transition probability is symmetric for forward and reverse moves.
Since $V_0 \supseteq V_{\mathrm{excl}}'\backslash V_{\mathrm{excl}}$, it follows that
\begin{equation}
\overline{V_0} \subseteq \overline{V_{\mathrm{excl}}'\backslash V_{\mathrm{excl}}} \subseteq \overline{V_{\mathrm{excl}}'} \cup V_{\mathrm{excl}} = V_f' \cup V_{\mathrm{excl}}
\end{equation}
and
therefore $\overline{V_0} = \overline{V_0} \cap (V_f' \cup V_{\mathrm{excl}})$.
Hence, applying the distributive law,
\begin{equation}
V_f \cap \overline{V_0} = V_f \cap \overline{V_0} \cap (V_f' \cup V_{\mathrm{excl}}) = V_f \cap \overline{V_0} \cap V_f',
\label{eq:VfinvV0}
\end{equation}
because $V_f \cap V_{\mathrm{excl}} = \emptyset$.

Using Eq.~\eqref{eq:VfinvV0} we rewrite the transition probability Eq.~\eqref{eq:transition_prob_V0} as
\begin{eqnarray}
\nonumber
\pi_{M\to M'} &=& e^{-\beta H_{cc}}\mathrm{min}\left(1,e^{-\beta \Delta H_{cc}}\right)
P^{\mbox{\tiny coll}}_{\mbox{\tiny trial}}(M\to M')\\
&&e^{z_p \left[\mu(V_f \cap V_f' \cap V_0) + \mu(V_f \cap V_f' \cap \overline{V_0})\right]},
\end{eqnarray}
and because the measures in the exponent are taken from disjoint sets we can simplify this equation as
\begin{eqnarray}
\pi_{M\to M'} &=& e^{-\beta H_{cc}}\mathrm{min}\left(1,e^{-\beta \Delta H_{cc}}\right)
P^{\mbox{\tiny coll}}_{\mbox{\tiny trial}}(M\to M')\nonumber\\
&&e^{z_p \mu(V_f \cap V_f')}
\label{eq:transition_prob_final}
\end{eqnarray}
This is the same transition rate as Eq.~\eqref{eq:detailed_bal_product},
consequently our restricted sampling algorithm obeys detailed balance.

We may choose $V_0$ as the smallest region with $V_0 \supseteq
V_{\mathrm{excl}}'\backslash V_{\mathrm{excl}}$ that is convenient to sample
from. E.g., we can sample in the excluded volume $V'_{\mathrm{excl},i}$ of the
single moved colloid $i$ at the position of the new configuration $M'$ {\em
only}, ignoring depletants that overlap with the colloid in the old
configuration $M$. For anisotropic colloids, we will choose the circumsphere of
diameter $d_{\mathrm{colloid}}+d_{\mathrm{depletant}}$ around the colloid in
the new configuration, as done in Fig.~\ref{fig:algorithm}.

We remark that a further possible optimization consists in restricting the sampling to the excluded volume
{\em shell} of the moved colloid $V_{\mathrm{excl},i}\backslash V'_{\mathrm{core},i}$, and it can be shown,
using steps analogous to above, that such a choice also fulfills detailed balance.

\subsection{Configurational bias moves}
\label{sec:configurational_bias}
The algorithm described above gives finite acceptance rates for translation
step sizes $\delta \lesssim z_p^{-1} R^{-2}$, where $R$ is the size of the colloid,
which is in general anisotropic. However, when there is more than
one depletant in the excluded volume shell around the colloid particle on average,
moves will be rejected most of the time. Equilibration of colloids in very dense depletant
systems is therefore difficult.

\begin{figure}
\includegraphics[width=\columnwidth]{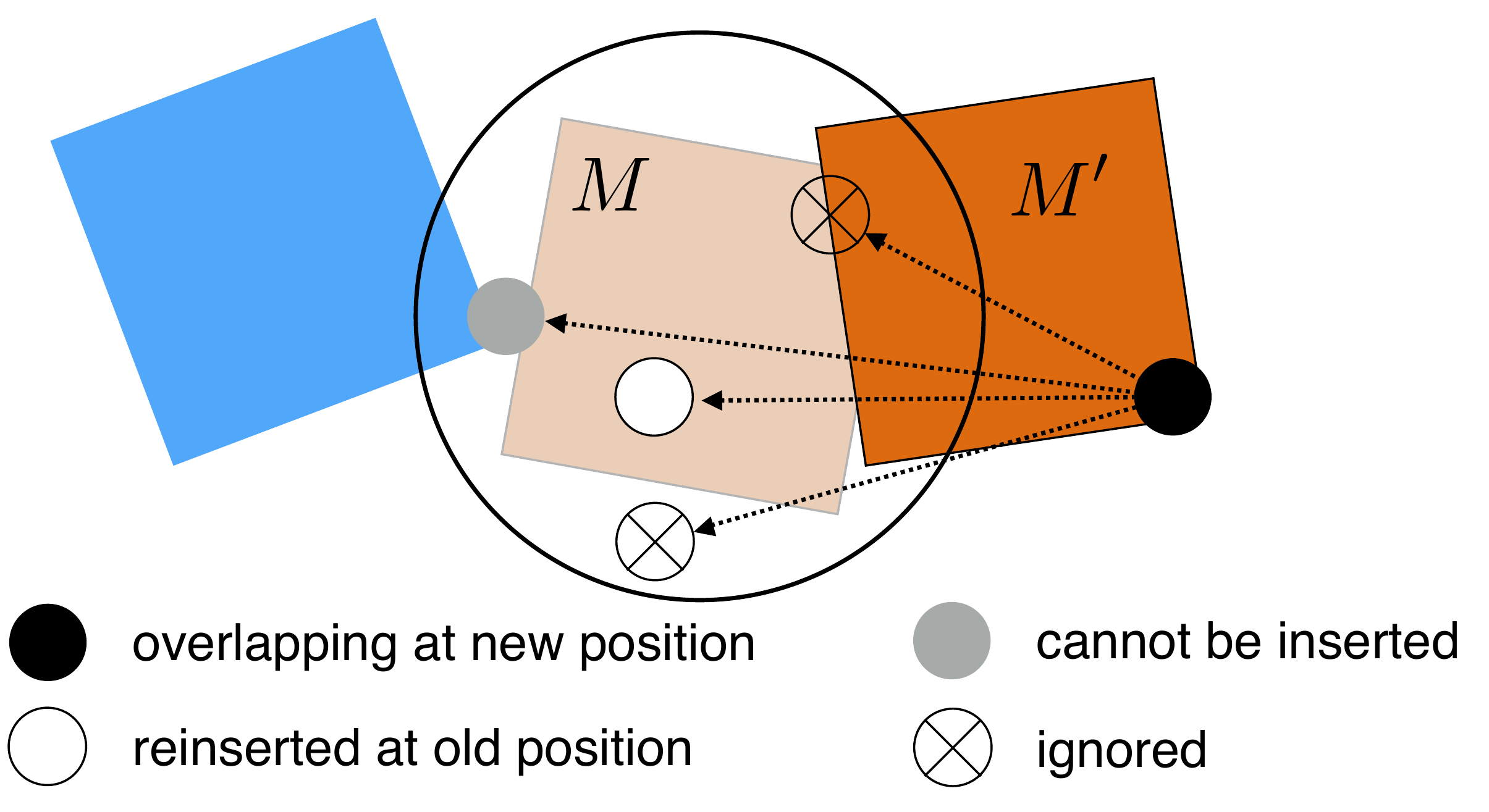}
\caption{Computation of the configurational bias weight for the forward move.
When a single moved colloid overlaps with a randomly inserted depletant in the
new configuration $M'$, we attempt to reinsert it $n_{\mathrm{trial}}$ times
such that it overlaps with the shape in the old configuration $M$. Valid
insertion attempts are those where the depletant neither overlaps with a
surrounding colloid nor with the colloid in the new position. The
configurational bias weight is computed from the number of successful
reinsertions, cf.\ Eq.~\eqref{eq:pacc_cb}.}

\label{fig:configurational_bias}
\end{figure}

To ameliorate this situation, we apply the configurational bias move of Biben,
Bolhuis and Frenkel \cite{Biben1996,Bolhuis1994} to implicit depletants, the
idea of which we briefly summarize. Figure~\ref{fig:configurational_bias}
depicts the basic idea.  For every depletant overlapping in the new
configuration $M'$, we attempt to reinsert it $n_{\mathrm{trial}}$ times such
that it overlaps with the shape in the old configuration $M$, but does not
overlap with any other colloid. Such a cluster move obeys detailed balance
because when performing the reverse move from $M'$ to $M$, the reinserted
colloid will overlap in the old configuration.  To correct for the
configurational bias generated in this way \cite{Siepmann1992}, we modify the
acceptance probability
\begin{equation} P_{\mbox{acc}} = \min\left(1,
\prod\limits^{N_{\mathrm{overlap}}}_{i=1}\frac{N'_{\mbox{\tiny insert,i}}
(N_i+1)}{(N_{\mbox{\tiny insert,i}}+1) N_i} \right),
\label{eq:pacc_cb}
\end{equation}
in which $N_{\mathrm{insert},i}$ and $N_{\mathrm{\tiny insert},i}'$ are the
number of times the overlapping depletant $i$ could be reinserted without
overlap into the old and new configuration, respectively.  The numbers $N_i,
N_i' \le n_{\mathrm{trial}}$, count the valid insertion attempts in which the
depletant overlaps with the moved shape in the old (new) configuration, without
overlapping in the other. All other insertion attempts are ignored.  The
increment of one ($N_{\mathrm{insert}}+1$) is necessary because the depletant
the colloid was overlapping with originally counts as a successful
reinsertion attempt for the reverse move.

\subsection{Parallel implementation}
\label{sec:parallel}
An important feature of our algorithm is that the depletant insertions are
independent and can be performed in parallel. We exploit this feature
to implement the algorithm on the GPU. Some details of the GPU implementation
are described in App.~\ref{app:gpu}.

In addition, depletants are inserted only in a local neighborhood of the
particle, reflecting the short-ranged nature of the depletion interaction. This
means the parallelization scheme for particle based Monte Carlo that has
recently been introduced within the Hard Particle Monte Carlo (HPMC) framework
\cite{Anderson2013,Anderson2015} in HOOMD-blue
\cite{Anderson2008,Glaser2015a,hoomd-url} can be generalized to our implicit
depletion algorithm. HPMC uses a checkerboard decomposition to allow
parallelization of the MC simulation on a graphics processor (GPU).  The
checkerboard is colored in such a way that simultaneously active cells are
separated by a layer of inactive cells of width
$d_{\mathrm{colloid}}+d_{\mathrm{depletant}}$, which allows the active cells to
be updated independently. Particles are not allowed to move outside their
cells. The checkerboard coloring is permuted randomly.  In order to
maintain ergodicity, the grid lines are randomly shifted.  HPMC also runs on
the CPU, using an efficient tree-based particle data storage for overlap checks
in combination with a sequential algorithm. Both the CPU and the GPU code path
can be combined with spatial domain decomposition \cite{Glaser2015a}, using the
same same concept of an inactive layer for parallel execution. A reference
implementation of the algorithm described in this paper will be
released open-source as part of HOOMD-blue\cite{hoomd-url}.

%\subsection{Minor optimizations}
%The local algorithm described above allows for some optimizations, such as
%only placing the depletants in excluded volume shell of the {\em in-sphere} of
%the colloid, instead of the entire circumsphere. These and some details
%of the GPU implementation are described in Sec.~\ref{app:optimizations}.

\section{Validation}
\label{sec:validation}
\subsection{Equation of state of the penetrable hard sphere model}

\begin{figure}
\includegraphics[width=\columnwidth]{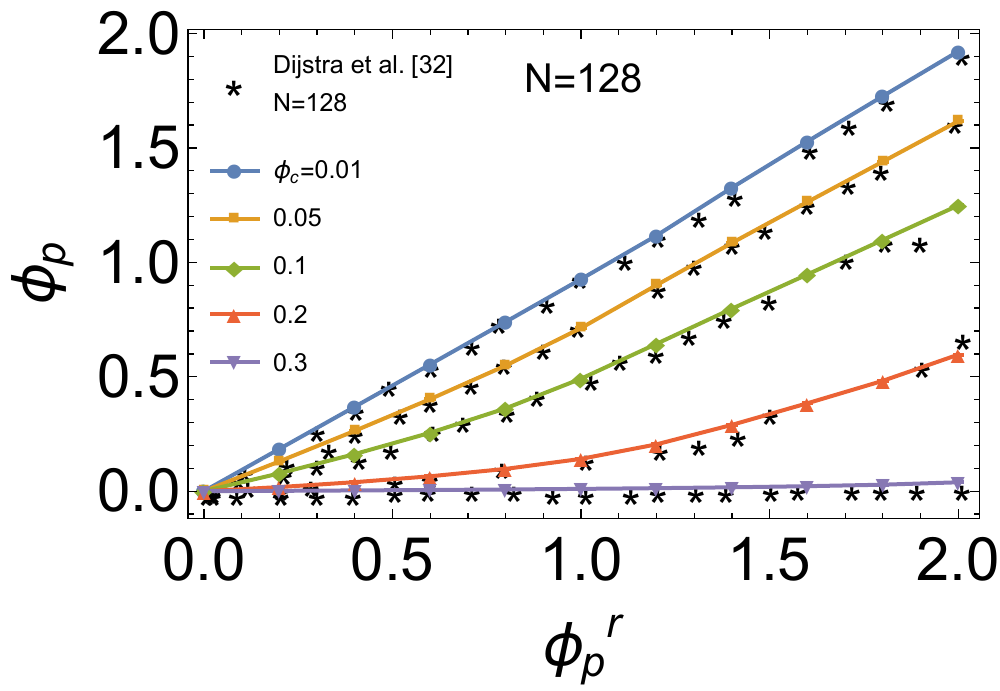}
\centering\includegraphics[width=\columnwidth]{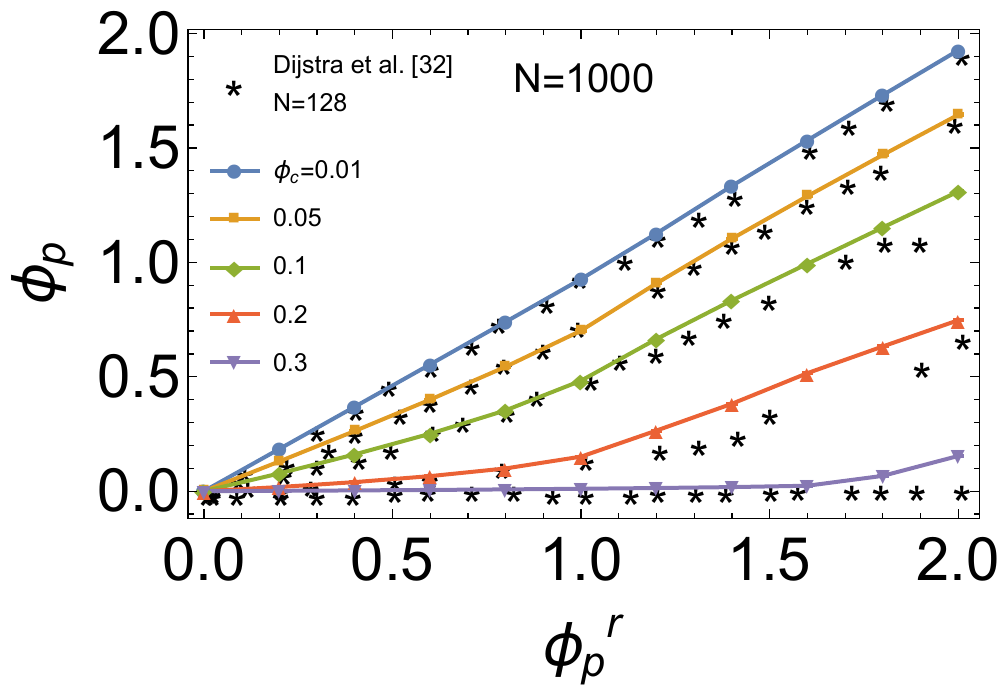}
\caption{Equation of state of spheres in penetrable hard sphere depletants. Plotted is the
measured free volume $\phi_p$ available to the penetrable hard spheres of
size ratio $q = d_{\mathrm{dep}}/ d_{\mathrm{colloid}} = 1$ vs. the reservoir volume fraction $\phi_p^r$ of the depletants,
for different hard sphere volume fractions $\phi_c = 0.01\dots 0.3$ (filled symbols).
Data by Dijkstra et al.~\cite{Dijkstra2006} for $N=128$ is shown as asterisks. {\em Upper panel}:
equation of state for $N=128$ colloids, {\em lower panel}: $N=1000$. The shown data includes
error bars taking into account only statistically independent samples \cite{Flyvbjerg1989a}.}
\label{fig:validation_small}
\end{figure}

To validate our method, we compare results for hard spheres with the
previously obtained results by Dijkstra et al.~\cite{Dijkstra2006}. We note
that even though their implicit algorithm for depletion does not obey detailed
balance, it relies on minimizing errors from the violation of detailed balance
through increasing the discretization of the MC integration step, which is a
trade-off between accuracy and performance. In order to obtain an accurate
equation of state, Dijkstra et.~al had to restrict themselves to fairly small
systems of $N=128$ spheres. Fig.~\ref{fig:validation_small} compares results
obtained with our algorithm (filled symbols) to those from Fig.~2 of
Ref.~\citenum{Dijkstra2006} (stars).  We show the measured free volume fraction
$\phi_p$ available to the penetrable hard spheres of same size, as a function of
the reservoir volume fraction $\phi_p^r$ for different colloid volume fractions
$\phi_c$ at constant simulation volume. For a system size of $N=128$ colloids,
our and Dijkstra's results are in essentially perfect agreement, mutually
validating both algorithms (top panel). However with our new algorithm we can
easily perform simulations for a larger system of $N=1000$ spheres. We do see
slight deviations from the results for the $N=128$ system (lower panel),
particularly at high depletant reservoir densities $\phi^r_p$, indicating the
presence of finite size effects for this system size.

\subsection{Coexistence curve of the penetrable hard sphere model}
We also tested the capability of our algorithm to equilibrate hard sphere
systems at gas-liquid coexistence, and especially near the critical point. We
carried out Gibbs ensemble simulations of hard spheres in penetrable hard sphere
depletants\cite{Panagiotopoulos1988}.  These types of simulations require
insertion of the colloid at random positions in the simulation box, which is
nearly impossible for high depletant fugacities. To overcome this difficulty,
we resort to the configurational bias scheme discussed in
Sec.~\ref{sec:configurational_bias} and originally introduced in the context of
the Gibbs ensemble of hard spheres with depleting rods in
Ref.~\citenum{Bolhuis1994}. For every exchange of a colloid between boxes,
depletants are randomly inserted at the new position, and overlapping
depletants are attempted to be reinserted in the old box.  The move is accepted
with the probability that accounts for the configurational bias weight.

\begin{figure}
\centering\includegraphics[width=\columnwidth]{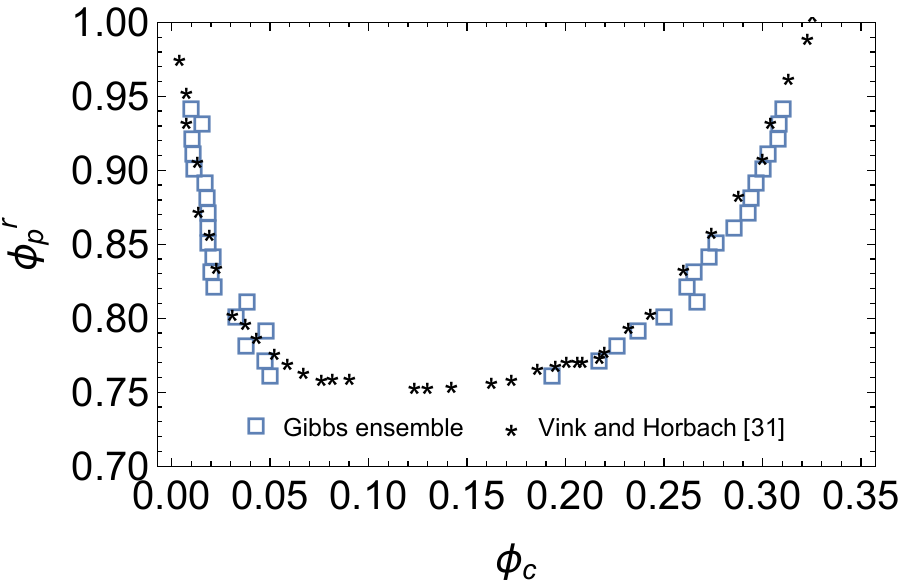}
\caption{Coexistence curve for phase separating hard spheres in the presence of
penetrable hard sphere depletants. Spheres ($N=512$) of initial packing
fraction $\phi_c=0.12$ are simulated in the semigrand Gibbs ensemble using at
constant normalized depletant reservoir density $\phi_p^r \equiv (\pi/6)
d_{\mathrm{dep}}^3 z_p$ using implicit depletants ($n_{\mathrm{trial}}=100$),
and the coexisting colloid volume fractions (squares) are obtained by fitting
the peaks of the two-dimensional $N-V$ histogram \cite{Smit1989a}. Asterisks
denote data from Ref.~\citenum{Vink2004} measured in the grand-canonical
ensemble.}

\label{fig:gibbs_ensemble}
\end{figure}

In Fig.~\ref{fig:gibbs_ensemble} we compare the coexistence curve thus obtained
to published data by Vink and Horbach \cite{Vink2004}.  Those authors did not
use the Gibbs ensemble, but performed direct simulation in the grand-canonical
ensemble of the colloids and depletants in a single box. Their method is
advantageous to sample the gas-liquid separation, which takes place at
intermediate densities $\phi_c \lesssim 0.4$, because it relies exclusively on
particle insertion and deletion at random positions in the simulation box.
Thus, in this regime their scheme can be at least as efficient as single
particle moves, if the particle deletions are combined with depletant
insertions, and vice versa.  However, the grand-canonical method is not easily
applicable to solid phases, for which particle insertion in a crystal lattice
is nearly impossible.  Our method, in contrast, computes depletion interactions
for single-particle translations and rotations.

As shown in Fig.~\ref{fig:gibbs_ensemble}, our data for the total system size
$N=256$, corresponding to the larger of the two system sizes studied by Vink
and Horbach, generally reproduces their data for a depletant-colloid size ratio
of $q=0.8$, at which many-body effects are important. However, we see some
scatter in our data, which is likely a consequence of surface effects that make
it notoriously hard to study coexistence near the critical point in Gibbs
ensemble simulations \cite{Smit1989a,Frenkel2001}. Vink and Horbach improved
their sampling using the umbrella method and thermodynamic integration.
Overall, however, our data obtained without using advanced free energy
techniques is in agreement with the published data, validating the method.

\section{Results}
\label{sec:results}

\subsection{Aggregation of hemispheres into superlattices}
Equilibrium data of anisotropic particles aggregating into crystals with
depletants is scarce \cite{Rossi2015}. Here, we present new results on the
hierarchical assembly of hemispheres into FCC/HCP-cluster phases.  Hard
hemispheres for self-assembly have been the subject of previous investigation.
Marechal and Dijkstra predicted the stability of a cluster-FCC (fcc$^2$) phase
for hemispheres, but they were unable to find it in self-assembly simulations
of sufficient size\cite{Marechal2010a}. Cinacchi presented the phase diagram of
hard spherical caps, which does not include an fcc$^2$
phase\cite{Cinacchi2013a}. Neither study involved depletants.

\begin{figure}
%\centering\includegraphics[width=\columnwidth]{HCP_hemispheres}
%\centering\includegraphics[width=\columnwidth]{hemispheres_q0_150_phi0_575_N512_etap0_400_ntrial0_snapT.png}
\includegraphics[width=\columnwidth]{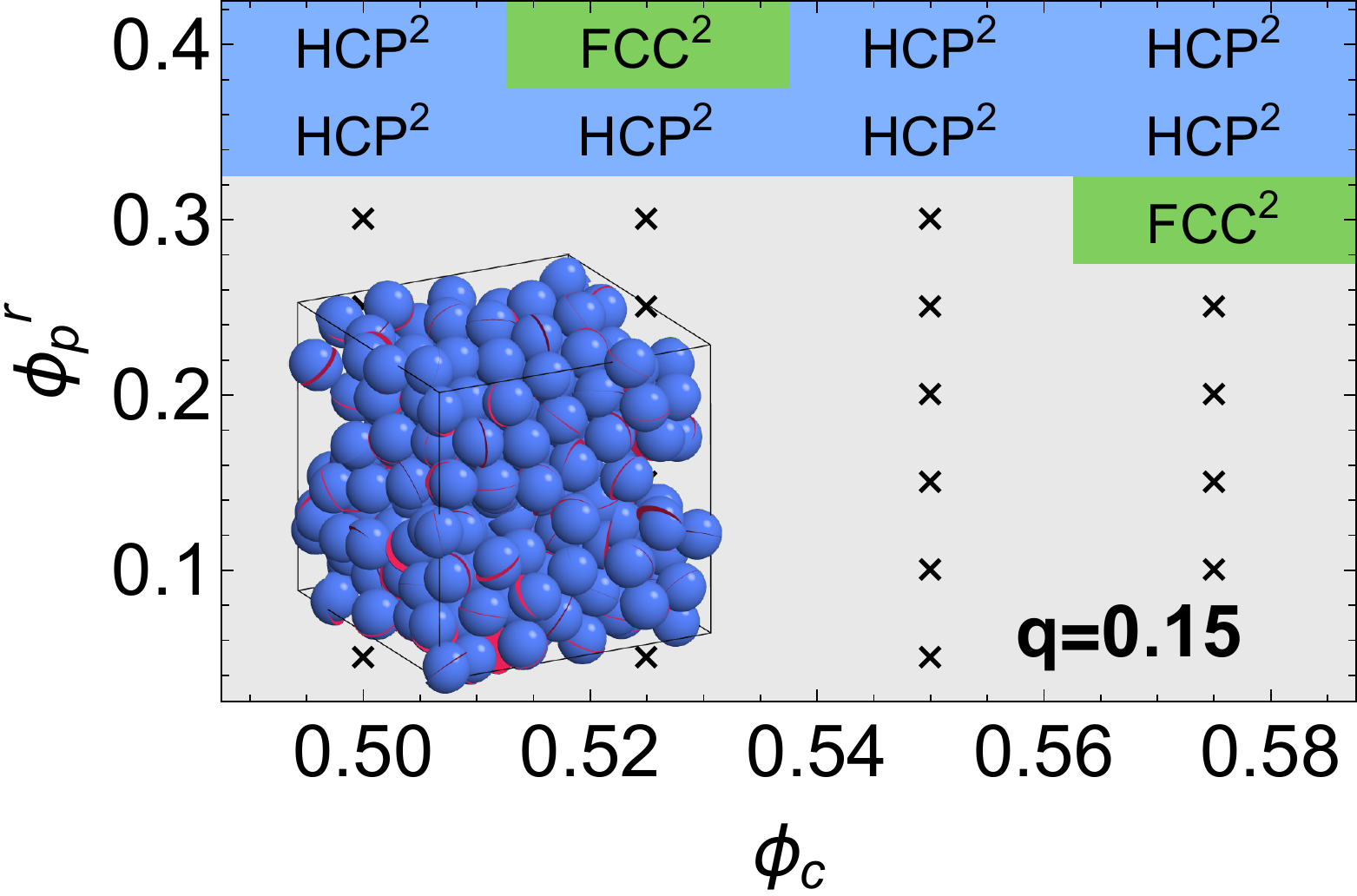}
\caption{Self-assembly of hemispheres into crystalline phases. Shown is the
kinetic phase diagram for $N=512$ hemispheres obtained with implicit simulation of
depletants as function of the depletant reservoir density $\phi_p^r$ and the colloid
density $\phi_c$, at depletant-hemisphere diameter ratio $q=0.15$. {\em Inset:}
Snapshot of the hcp$^2$ phase found for $\phi_c=0.575$ and $\phi_p^r=0.4$. Similar
phase diagrams were obtained for $q=0.175$ and $q=0.125$ (not shown).}
\label{fig:hemispheres_q015}
\end{figure}

We analyze the phase behavior of hemispheres in the presence of penetrable hard
sphere depletants.  Figure \ref{fig:hemispheres_q015} shows the kinetic phase
diagram as a function of depletant reservoir density $\phi_p^r$ and colloid density
$\phi_c$, for a depletant-hemisphere diameter ratio of $q=0.15$.  Remarkably,
we observe the formation of the fcc$^2$ and hcp$^2$ phases at finite
depletant densities $\phi_p^r \ge 0.30$, and the inset shows a snapshot of such a
configuration of hemispheres. However, at zero depletant fugacity, which
corresponds to the case studied previously, we did not observe any ordered
phase, even after $6\times10^8$ MC sweeps. Instead, we find a cluster fluid.
In the phase diagram, we find close-packed crystals with both HCP and FCC stacking,
and we suspect the fact that both occur indicates that the free energy difference is small
\cite{Frenkel1984a}.

\begin{figure}
\centering\includegraphics[width=\columnwidth]{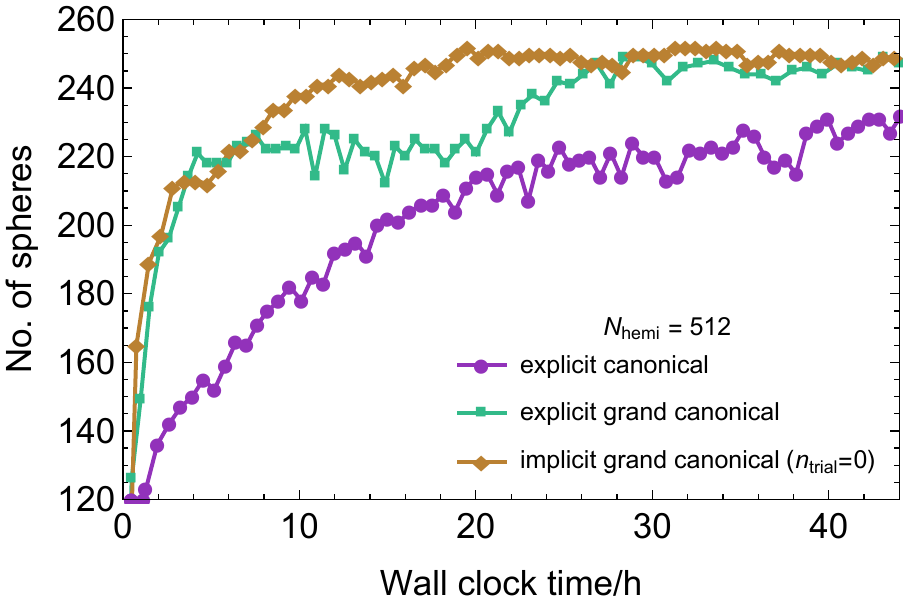}
\caption{Aggregation kinetics of hemispheres. Shown is the number of spheres
formed after simulation time $t$ (in hours), for a simulation with implicit
depletants (diamonds), explicit grand-canonical depletants (squares) and
canonical depletants (circles). Simulations where performed at colloid volume
fraction $\phi_c=0.575$ and depletant reservoir density $\phi_p^r=0.40$ for a
depletant-colloid diameter ratio of $q=0.175$, on eight cores of an Intel Xeon
E5-2680 processor with spatial domain decomposition via MPI (single precision).
A sphere is defined as two hemispheres with their face centers being closer
than $0.2 d$ apart, where $d$ is the diameter of the (hemi-)sphere. In the
canonical case, the constant number $N_p=4884$ of explicit depletant particles
has been chosen to be the average number of depletants in the free volume of
the grand-canonical simulations, after phase transformation.}
\label{fig:aggregation_hemispheres}
\end{figure}

We compare the implicit method against two other schemes, an explicit
grand-canonical ensemble for the depletants \cite{Frenkel2001}, and a canonical
ensemble with fixed concentration of depletants.  Figure
\ref{fig:aggregation_hemispheres} shows the number of hemisphere pairs that
have formed after time $t$. Because Monte Carlo simulations do not have a time
scale, we choose the wall-clock time of the simulation as an ad-hoc measure of
time.  By analyzing bond order, we found that the time scale of crystallization
corresponds to the time when all 512 hemispheres in the simulation box have
paired up. This event occurs earliest for the implicit depletion algorithm.
The simulation with explicit grand-canonical depletants also orders at a
later time. However, the simulation with fixed number of depletants does not
equilibrate into an ordered phase within the wall-clock time limit of 48h or
$7.8\times10^7$ sweeps. Our findings show that the implicit algorithm leads to
the fastest assembly of hemispheres into cluster crystal phases.

\subsection{Diffusivity of discoids with depletants}
Ellipsoids are simple examples of anisotropic particles. Recently, discoids have been demonstrated
to arrange into metastable strand structures at sufficiently high density of polymeric
depletants \cite{Hsiao2015a}. Here, we investigate the diffusivity of discoids at depletant
densities that do not lead to ordering. For Monte Carlo simulations with single particle moves,
the diffusivity of the colloids in terms of mean square displacement per wall clock time is
an effective measure of the speed of equilibration of the simulation. In our simulations,
we tune the single particle step size for translation and rotation so as to yield an average
acceptance rate of $20\%$.

\begin{figure}
\centering\includegraphics[width=\columnwidth]{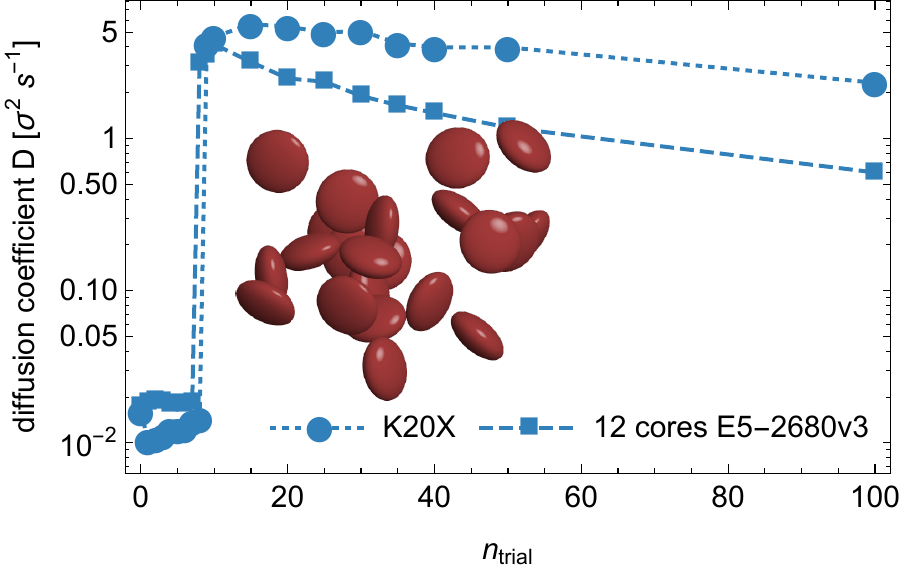}
\includegraphics[width=\columnwidth]{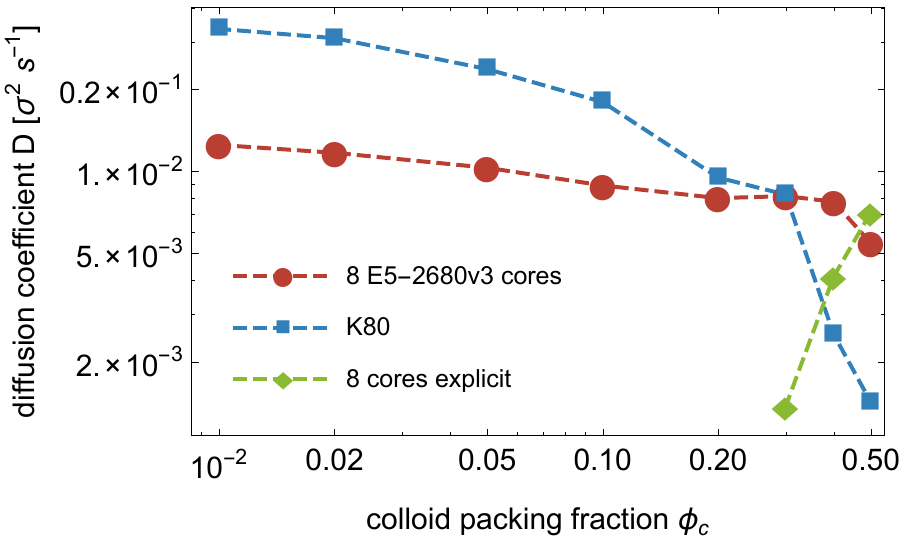}
\caption{Diffusivity of discoids in penetrable hard sphere depletants.  {\em
Upper panel:} Diffusion coefficient vs. the number $n_{\mathrm{trial}}$ of
configurational bias swaps, for a simulation of $N=500$ discoids on 12 CPU
cores (Intel Xeon E5-2680v3) using MPI (squares) and a single NVIDIA K20X GPU
(circles). For simulation parameters, see main text.  The diffusivity is
obtained from fitting the linear mean square displacement MSD as function of
the wall-clock time $t$ (in seconds).  {\em Lower panel:} Diffusion coefficient
vs. colloid density $\phi_c$ ($n_{\mathrm{trial}} = 0$), for a simulation on a
NVIDIA K80 GPU (squares), on 8 cores of an Intel Xeon E5-2680v3 (circles), and
for a simulation of explicit depletants in the grand-canonical ensemble
(diamonds), on the same hardware.}
\label{fig:bench_ellipsoids}
\end{figure}

The upper panel of Figure \ref{fig:bench_ellipsoids} shows the effect of
$n_{\mathrm{trial}}$ on the diffusivity $D$ of discoids. The colloid particles
are uniaxial ellipsoids with semi axes $a=b=0.5$ and $c=0.25$, the depletants
are of radius $r=0.25$, and the simulations are performed in a dilute system at
colloid density $\phi_c=0.01$ and depletant reservoir density $\phi_p^r=0.40$,
below the coexistence density for metastable clusters \cite{Hsiao2015a}. From
the graph, it can be clearly seen that using configurational bias moves with a
modest value of $n_{\mathrm{trial}} \gtrsim 10$ speeds up the equilibration by
almost three orders of magnitude compared to not using configurational bias
moves. The effect is dramatic and similar in magnitude between running the
simulation on the CPU vs. the GPU. At peak diffusivity, there is a slight
advantage to using the GPU, compared to CPU socket performance. For higher
values of $n_{\mathrm{trial}}$, the performance drops off slowly, as a result
of the increased computational effort to carry out the depletant reinsertions,
while the effect of increasing the step size due to a higher acceptance ratio
is weaker. We note that we carried out simulations with finite values of
$n_{\mathrm{trial}}$ at higher colloid densities as well (data not shown) and
found the effect to be less pronounced at these densities.

We further measure the performance at different colloid densities $\phi_c$
between the dilute regime and the regime of a dense liquid, for the same
parameters as above, with $n_{\mathrm{trial}}=0$
(Fig.~\ref{fig:bench_ellipsoids}, lower panel). For simulations with implicit
depletants, either using the CPU or the GPU, the performance depends only
slightly on the colloid volume fraction, directly confirming the beneficial
effect of implicit calculation of the interaction in the dilute
system, where the number of depletants would be very high with an explicit
treatment.  Indeed, the performance of the explicit depletant simulations in
the grand-canonical ensemble drops noticeably when going from $\phi_c=0.50$
towards lower densities, and the system becomes practically impossible to
equilibrate when $\phi_c < 0.30$. Looking at GPU vs. CPU performance, we note
that GPUs are advantageous for very dilute systems, but do not provide better
performance when the system is dense in colloids. This is because the
checkerboard parallelization scheme implemented for performing the colloid
moves on the GPU (Sec.~\ref{sec:parallel} and Ref.~\citenum{Anderson2015})
requires a large simulation box to operate efficiently.

\section{Conclusion}
\label{sec:conclusion}

We have presented an efficient algorithm to implicitly simulate depletion
interactions between anisotropic colloids. The algorithm is implemented on
parallel multi-core processors and graphics processing units. Combined with a
parallel Monte Carlo scheme \cite{Anderson2013,Anderson2015}, the algorithm
offers a way to tackle large scale simulations of hard shapes with depletants.
The scheme may be readily generalized to soft interactions between the colloid
and the depletant, such as the Hertz potential \cite{Rovigatti2015}.  We stress
that even though the algorithm is parallel, already its serial implementation
offers significant speed-ups over algorithms that do not use cluster moves, for
dilute systems of colloids, because only depletants in the neighborhood of
every particle are considered.  Nevertheless, the method works perfectly well
for the fluid-solid transition.

We see applications for our method in the simulation of anisotropic colloid
phase behavior.  Even without depletants, polyhedra have been shown to order
into a multitude of different structures \cite{Damasceno2012}. With depletion
interactions, additional phases can be
stabilized\cite{Henzie2012,Young2013a,Rossi2015,Karas2015a}. The algorithm can
also be used to study the aggregation of entropically patchy colloids into
colloidal polymer chains, held together by strong depletion bonds
\cite{Ashton2015}. In this context, it would be interesting to study solutions
as well as melts of such colloidal polymers. An interesting open question
concerns whether depletant entropy can stabilize not only close-packed but also
open ordered structures \cite{Mao2013}. In protein crystallization, depletant
polymers are commonly used as precipitants. An important limitation of our
algorithm is that it treats only non-interacting depletants, and the validity
of that approximation remains to be investigated for specific systems. In
contrast to enthalpically patchy models, our algorithm does not require
implementation of shape-specific attractive patches to study aggregation of
colloids, and the algorithm is therefore highly robust and generic.

\begin{acknowledgments}
We are thankful to Werner Krauth for a discussion that led to the development of this
algorithm. We also thank Michael Engel for fruitful discussions and careful reading of
the manuscript.

This material is based upon work supported in part by the U.S. Army Research
Office under Grant Award No. W911NF-10-1-0518 and by a Simons Investigator
award from the Simons Foundation to Sharon Glotzer.  This research used the
Extreme Science and Engineering Discovery Environment \cite{Towns2014} (XSEDE),
which is supported by National Science Foundation grant number ACI-1053575;
XSEDE award DMR 140129. The Glotzer Group at the University of Michigan is an
NVIDIA GPU Research Center. Hardware support by NVIDIA Corp. is gratefully
acknowledged.
\end{acknowledgments}

\begin{appendix}
\section{GPU implementation}
\label{app:gpu}

In the GPU implementation, we perform the colloid trial moves in the active
cells\cite{Anderson2015} and the depletant insertions in different kernels. To
insert depletants, we draw a random number of depletants for every moved
colloid, as described in Sec.~\ref{sec:basic}. We use a one-to-one mapping
between depletants and thread groups of size $n\le n_{\mathrm{max}}$. Here,
$n_{\mathrm{max}}=32$ is the maximum number of threads that can perform overlap
checks synchronously, and we tune $n$ at run-time. When any thread detects an
overlap between the depletant and any particles in the old configuration, the
depletant is ignored. In the other case, if the depletant overlaps with the
moved colloid, that colloid move is flagged for rejection.

When the configurational bias scheme is used ($n_{\mathrm{trial}} > 0$), a
second kernel with a similar thread mapping is launched, however, depletants
are assigned to whole thread blocks of size $s \le 1024$, which is an
auto-tuned parameter, so that the bias weights of different reinsertions
belonging to the same depletant can be summed in shared memory.

%\section{Faceted sphere overlap check}
\end{appendix}
\end{document}